# Spatial beam self-cleaning in multimode lanthanum aluminum silicate glass fiber

Romain Guénard[1], Katarzyna Krupa[2], Alessandro Tonello[1], Marc Fabert[1], Jean Louis Auguste[1], Georges Humbert[1], Stéphanie Leparmentier[1], Jean-René Duclère[6], Sébastien Chenu[6], Gaëlle Delaizir[6], Guy Millot[2], Daniele Modotto[3], Stefan Wabnitz[4,5] Vincent Couderc[1]

[1]*Université de Limoges, XLIM, UMR CNRS 7252, 123 Avenue Albert Thomas, 87060 Limoges, France*
[2]*Université Bourgogne Franche-Comté, ICB, UMR CNRS 6303, 9 Av. A. Savary, 21078 Dijon, France*
[3]*Dipartimento di Ingegneria dell'Informazione, Università di Brescia, via Branze 38, 25123 Brescia, Italy*
[4]*Dipartimento di Ingegneria dell'Informazione, Elettronica e Telecomunicazioni, Sapienza Università di Roma, via Eudossiana 18, 00184 Rome, Italy*
[5]*Istituto Nazionale di Ottica del Consiglio Nazionale delle Ricerche (INO-CNR), Via Campi Flegrei 34, I-80078 Pozzuoli (NA), Italy*
[6] *Institut de Recherche sur les Céramiques, UMR 7315 CNRS-Université de Limoges, Centre Européen de la Céramique, 12, rue Atlantis, 87068 Limoges Cedex, France*

*Corresponding author:* Vincent.couderc@xlim.fr

We demonstrated that spatial Kerr beam self-cleaning can be obtained in a highly multimode multicomponent optical fiber based on lanthanum aluminum silicate oxide glasses ($SiO_2$-$Al_2O_3$-$La_2O_3$), which was made by using the modified powder in tube technology (MIPT). We show how such fabrication method can provide interesting potentialities to design doped multimode optical fibers with homogeneous and quasi-parabolic refractive-index core profile for nonlinear optics applications.

## 1. Introduction

Graded-index multimode optical fibers (GRIN MMFs) have been recently reconsidered, owing to their potential to implement complex nonlinear spatiotemporal phenomena, involving a rich variety of novel nonlinear processes [1-15]. Among others, we may list, for instance, Kerr-beam self-cleaning (which allows for brightness improvement of a multimode beam) [4], geometric parametric instability (GPI) [5-6], multimode solitons [7], intermodal four-wave mixing (IMFWM) [8-10], modulation instability [11], ultrabroadband dispersive radiation generation [12], second harmonic generation induced by an all-optical poling process [13-14], supercontinuum generation [15-16], spatio-temporal mode-locking [17-18], and control of nonlinear multimode propagation by means of wavefront shaping [19-20].

Spatial Kerr-beam self-cleaning is of particular interest since it leads to force light to self-organize and to emerge at the fiber output end largely confined in a quasi-single transverse mode. The nonlinear mechanism at the origin of such peculiar spatial reshaping can be understood as a parametric four-wave mixing process, which breaks the orthogonality among the modes, and promotes an energy transfer towards the fundamental mode or, more generally, lower-order modes, even in the presence of linear mode coupling owing to fiber stress or bending. Indeed, the phase-matching process of this parametric process is initiated by the longitudinal beating between a multitude of modes, producing a periodic refractive index grating mediated by the Kerr effect. Because of the strong convergence of the beam energy from high-order modes (HOMs) towards a single transverse mode of the fiber, self-phase modulation comes into play, which prohibits the backward energy flow process into HOMs, thus irreversibly locking the beam in a quasi-single transverse mode.

In this self-organization process, one of the most important parameters to consider is the refractive index profile. In the case of a parabolic transverse index distribution, the guided modes can be subdivided into groups with equally spaced propagation constants. Such a parabolic index profile has been widely used since many years to minimize modal dispersion of MMFs, and it is particularly helpful to increase the nonlinear interaction length among modes. The Kerr-induced grating is at the origin of the aforementioned phase matching process, which is able to couple modes that, at low powers, have different propagation constants, hence permitting to control the Kerr-beam self-cleaning dynamics [4].

Spatial beam self-cleaning has also been reported in active multimode optical fibers even in the case of a non-parabolic refractive-index profile [17-18, 22]. However, our experiments show that a deviation from a perfect parabolic shape increases the power threshold for beam cleaning, and consequently reduces the amount of power transferred to the "cleaned" mode [23]. Therefore, besides the choice of the input conditions, the principal ingredients for beam self-cleaning are the fiber characteristics, including the refractive index profile, the core diameter and the nonlinear refractive index of the glass. In this context, fiber technology can play a pivotal role, by permitting to design complex guiding profiles with longitudinal and/or transverse structuration of dopants that can in turn enable the control of velocity differences among the guided modes, hence the efficiency of energy exchange among them. Unfortunately, the control of the refractive index profile and active ions currently remains a challenging task, which is not yet completely mastered.

In 1995 Ballato *et al.* [24] used a new fiber fabrication method called "powder in tube" with the aim of increasing rare-earth concentration in optical fiber lasers. This method was based on the melting of a powder composition in a capillary tube during the drawing process. It offers a large degree of freedom in the fiber design, thus allowing one to realize symmetric or asymmetric fibers composed of several different materials. Moreover, this method can be easily combined with the stack-and-draw process. With the powder in tube method, the mutual diffusion of different chemical compositions between the core and the cladding can contribute to shaping the refractive index profile and the doping concentration. Interestingly, such a technique may permit the fabrication of very large core active fibers with a controlled repartition of rare-earth ions. The method has been further improved by implementing a modified powder-in-tube technique (MPIT), which consists of a consolidation process taking place in the preparation of the preform, either under vacuum or pressurized gas, and where particular properties of the fiber were achieved [25].

In this paper, we demonstrate that by using the MPIT method, it is possible to draw a multimode lanthanum aluminum silicate fiber (MLASF) with quasi-parabolic refractive index profile, exhibiting losses that are low enough to permit observing nonlinear effects such as the spatial self-cleaning. Spatial, spectral and temporal dynamics of the output cleaned beam are also analyzed and discussed.

## 2. Fiber characteristics

In our experiment, we used a 3m-long optical fiber, fabricated by using the MPIT method, and composed by lanthano-aluminosilicate (SAL) glass with the following composition of the fiber core (in mol%): 82% $SiO_2$ -11.6 % $Al_2O_3$ - 6.4 % $La_2O_3$ [27]. Because of the diffusion species from the molten core to the silica cladding during the drawing process, a quasi-parabolic refractive index profile was obtained after the fiber drawing [25, 26] (see Fig.1). This glass has a transition temperature $T_g \sim 900$ °C and a thermal expansion coefficient $\alpha \sim 2.8 \times 10^{-6}$ $K^{-1}$: note that this value is 7 times larger than the corresponding value for pure silica glass. The refractive index difference between the core and the cladding is close to 0.06 for a core diameter of 28 μm. Thus, more than one thousand modes can be guided in such multicomponent optical fiber. The measured linear losses are 1.95 dB/m, and the nonlinear refractive index $n_2$ is estimated to be between 5.8 and 7 $\times$ $10^{-20}$ m² / W [27], which is 2 times higher than the one of the silica cladding $n_{2(silica)}=2.7 \times 10^{-20}$ m² / W. The zero dispersion wavelength of the fundamental mode at the pump wavelength is close to 1400 nm.

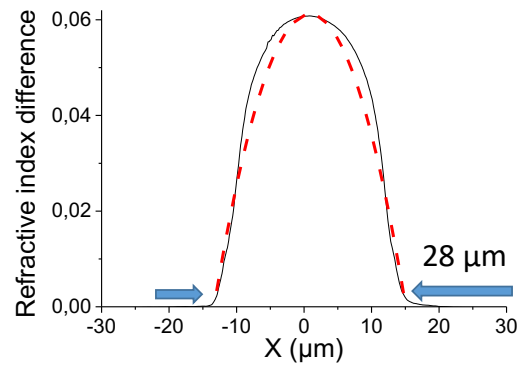

**Figure 1:** Refractive index difference core profile of the MLASF (black line); Parabolic fitting (red dashed curve).

## 3. Experimental setup

In our experimental setup, we used a longitudinally and transversely single-mode laser at 1064 nm, delivering more than 180 kW of peak power with pulses of 500 ps (repetition rate 500 Hz). An optical system, composed by a polarizer cube interposed between two half-wave plates, and a convergent lens, allowed to control the power coupled in the fiber and the orientation of the linear polarization state of the input light. The laser beam focused on the input fiber face was close to 18μm (Full Width at 1/e² in intensity). The output beam profile was measured with a CCD camera while the spectrum was recorded with an Ando AQ-6315A optical spectrum analyzer. The output temporal pulse shape was measured instead by using a 20 GHz bandwidth oscilloscope coupled with a 12 GHz photodiode.

## 4. Results and discussion

We started our experiments by characterizing the output spatial beam shape as a function of the power injected into the MLASF. As can be seen from the insets of Fig. 2, for low values of input peak power and for a centered input beam, the output beam exhibited a speckled pattern, owing to the simultaneous excitation of many transverse modes propagating with their own phase velocity. Despite the careful alignment of the input beam with the fiber axis, the beam quality at the output end of the fiber appeared to be quite poor, without the prevalence of any given transverse guided mode.

However, by increasing the input peak power and by keeping unchanged the position of the input beam, we observed a gradual change in the energy distribution that was progressively concentrated at the center of the fiber core. When we reached the maximum value of 32 kW peak power coupled in the multimode fiber, the diameter of the output beam reached a value close to 12 µm (at $1/e^2$). In the presence of this peak power variation, the $M^2$ coefficient, recorded with a Thorlabs M2MS-BC106VIS(/M) system, evolved from 4.5 down to 1.6, as illustrated in Fig. 2 and 3, respectively. The nonlinear beam reshaping observed at the fiber end shows a significant brightness enhancement of the propagating beam.

The peak power threshold of the Kerr self-cleaning effect was measured to be equal to 4 kW for a 3-m long optical fiber. For the sake of comparison, the threshold for spatial Kerr self-cleaning obtained in a 3-m long standard, commercially available, 50/125 GRIN optical fiber was close to 12 kW. Note however that, for a fair comparison between the two fibers, one should consider that the GRIN fiber has a core area which is 3.2 times larger than the present multicomponent fiber. Once rescaled by the area, the relatively low value of self-cleaning threshold, obtained for our multicomponent fiber, appears to be comparable with that obtained in a standard GRIN fiber, despite the much higher loss coefficient (1.95 dB/m for our fiber compared to 2.3 dB/km for GRIN fiber), but thanks to the twice higher nonlinear refractive index. Indeed, the higher nonlinear refractive index of the core material appears to balance, upon propagation, the faster input power decrease, so that similar Kerr self-cleaning thresholds are obtained with the MLASF and the GRIN fiber.

Next, we studied the spectral evolution upon input power that accompanies Kerr beam self-cleaning. As can be seen in Fig.4, for input peak powers ranging from 0 to 30 kW, the output spectrum of the multimode beam at 1064 nm changes significantly. In the linear propagation regime (for relatively low powers, i.e., 0.2 kW in Fig.4), the spectral profile exhibits a single longitudinal mode with spectral width close to 10 pm.

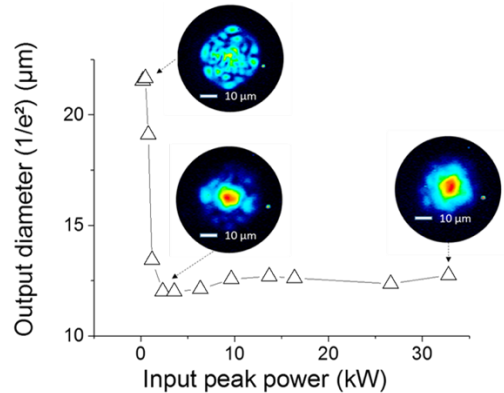

**Figure 2:** Output beam diameter (at $1/e^2$) as a function of the coupled peak power in the MLASF.

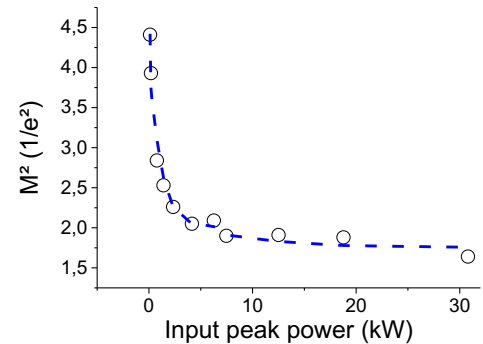

**Figure 3:** $M^2$ coefficient evolution measured at $1/e^2$ in intensity as a function of the coupled peak power in the MLASF.

When we increase the power close to the self-cleaning threshold (see curve for 3.2 kW in Fig.4), we can see that the self-cleaning process takes place with only minimal impact on the output spectrum profile: small pedestals appear at -35 dB below the maximum peak intensity. On the other hand, for peak powers higher than 10 kW, one observes a strong modification in the spectral power distribution. A modulation instability (MI) is clearly visible in the spectral profile for the 30 kW curve in Fig.4: a pair of sidebands grows at the edges of the spectrum. For the peak power of 30 kW, the spectral broadening of the central frequency reaches the Raman shift of lanthanum aluminum silicate oxide glass, leading to two lateral bands at 1039 nm and 1091 nm [29]. We also verified in these experiments that, in agreement to previously reported results [4], Kerr-beam self-cleaning appears before the occurrence of other nonlinear effects such as self-phase modulation, MI and stimulated

Raman scattering (SRS). The maximum spectral shift of the sidebands is close to 7.15 THz in the normal dispersion regime (see Fig.4).

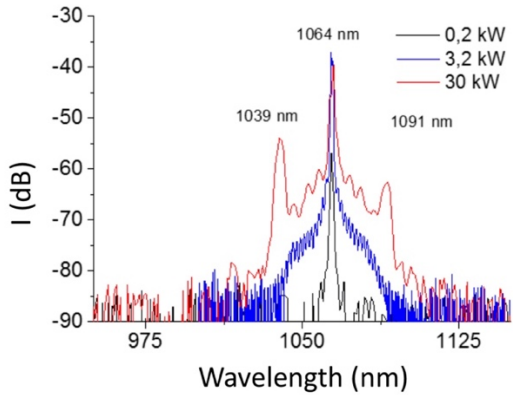

**Figure 4:** Output Spectrum at different input peak powers (fiber length: 3 m).

We completed our investigations by measuring the signature of the Kerr beam self-cleaning process in the time domain. In order to do so, we magnified the output beam, so that only the central part was detected by the small effective area of an ultrafast photodetector.

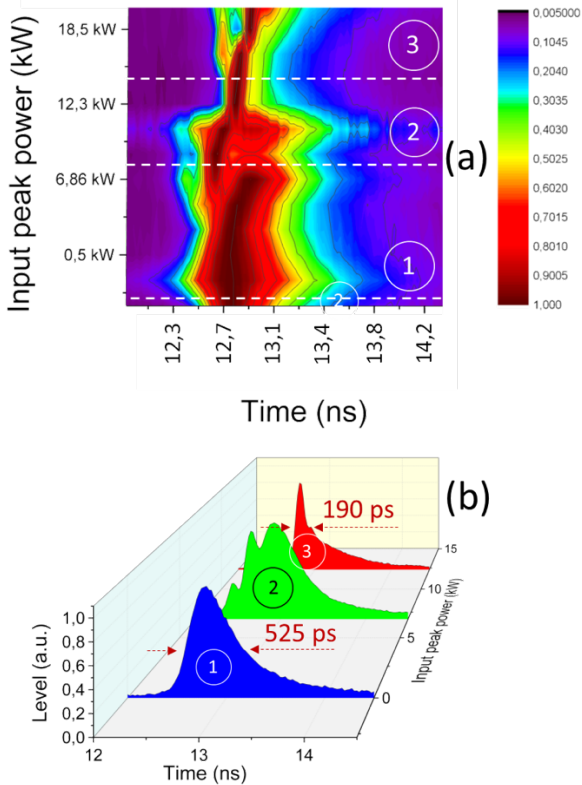

**Figure 5:** (a) Temporal evolution of the output pulse profile versus the input peak power; (b) Three different sample cases of pulse profiles for 0.01 kW, 4.3 kW, 12.3 kW (fiber length : 3m).

We recorded the temporal pulse profile at the center of the output beam, which mainly contains the fundamental transverse mode. We repeated the same operation for increasing input power values. Our results are summarized in Fig. 5. Because of the symmetry of the initial coupling conditions, where the propagation direction of the incident Gaussian beam was aligned along the fiber axis, we expect that the initial phases of each transverse mode are similar at the fiber input face.

By assuming a perfect phase-matching of the spatial four-wave mixing process, energy exchange among modes is mainly driven by the instantaneous peak power carried in each of them [30]. Therefore, an initial bell-shaped pulse is drastically modulated in time upon propagation, with the appearance of several sharp peaks for different values of the instantaneous intensity (see Figs. 5(a) and 5(b)). Such oscillating behavior (with respect to power variations) of the mode coupling process was observed in the central part of the output beam pattern.

In Fig. 6 we plotted the maximum peak power enhancement at the fiber output, obtained by taking into account ~~of~~ the pulse shape and its temporal duration. An enhancement factor of the initial peak power of more than 1.2 was measured in the presence of the Kerr self-cleaning process, just before the appearance of SRS. As soon as SRS is observed, a drastic depletion of the central part of the pulse occurs (see Fig.6), as can be seen for input peak powers higher than 15 kW.

In this way, we could demonstrate that beam self-cleaning is accompanied by a temporal pulse narrowing from 525 ps down to 190 ps (FWHMI) (see Fig. 5(b)). Similar results were obtained with a GRIN fiber [30].

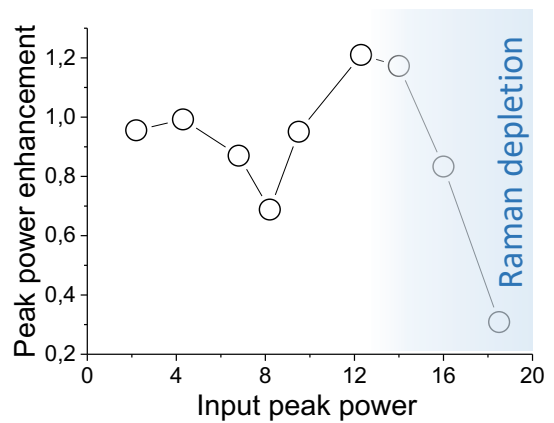

**Figure 6:** Peak power enhancement of the pulse peak recorded at the center of the output beam pattern (fiber length: 3 m; initial pulse width: 525 ps, compressed pulse duration for 12.3 kW: 190 ps)

## 5. Conclusion

In this paper, we demonstrated that the spatiotemporal dynamics associated to Kerr beam self-cleaning could be obtained in an "exotic" quasi-parabolic index lanthanum aluminum silicate multimode fiber drawn by using the MPIT method, where the propagation losses of the fiber were compensated by a higher nonlinear refractive index. Indeed, we observed a significant improvement of the output beam brightness, as well as a pulse shortening from 525 ps down to 190 ps with a limited distortion of the output spectrum profile.

The peak power enhancement obtained during the Kerr self-cleaning phenomenon is sufficient to reach the SRS threshold. Moreover, in this work we showed that MPIT is an excellent fabrication method to draw fibers for investigating and exploiting the spatiotemporal dynamics of Kerr beam self-cleaning, since it allows obtaining a graded-index profile of the fiber core composed of materials beyond the scope of the usual silica including, in particular, rare earth-ion compounds.

## Acknowledgements

This work was supported by the European Research Council (ERC) under the European Union's Horizon 2020 research and innovation programme (project STEMS grant No. 740355). This work benefited also of a financial support from the French state, managed by the National Agency of Research via the TRAFIC project (ANR-18-CE08-0016-01).

## References

[1] A. Picozzi, G. Millot, and S. Wabnitz, "Nonlinear virtues of multimode fibre," Nat. Photonics 9, 289–291 (2015).
[2] L. G. Wright, D. N. Christodoulides, and F. W. Wise, "Controllable spatiotemporal nonlinear effects in multimode fibres," Nat. Photonics 9, 306–310 (2015).
[3] S. Wabnitz, K. Krupa, V. Couderc, D. Modotto, A. Barthélémy, and G. Millot, "Nonlinear dynamics in multimode optical fibers," Proc. SPIE 10540, 105402B (2018).
[4] K. Krupa, A. Tonello, B. M. Shalaby, M. Fabert, A. Barthélémy, G. Millot, S. Wabnitz, and V. Couderc, "Spatial beam self-cleaning in multimode fibres," Nat. Photonics 11, 237–241 (2017).
[5] K. Krupa, A. Tonello, A. Barthélémy, V. Couderc, B. M. Shalaby, A. Bendahmane, G. Millot, and S. Wabnitz, "Observation of geometric parametric instability induced by the periodic spatial self-imaging of multimode waves," Phys. Rev. Lett. 116, 183901 (2016).
[6] C. Mas Arabí, A. Kudlinski, A. Mussot, and M. Conforti, "Geometric parametric instability in periodically modulated graded-index multimode fibers," Phys. Rev. A 97, 023803 (2018).
[7] W. H. Renninger and F. W. Wise, "Optical solitons in graded-index multimode fibres," Nat. Commun. 4, 1719 (2012).
[8] E. Nazemosadat, H. Pourbeyram, and A. Mafi, "Phase matching for spontaneous frequency conversion via four-wave mixing in graded index multimode optical fibers," J. Opt. Soc. Am. B 33, 144–150 (2016).
[9] R. Dupiol, A. Bendahmane, K. Krupa, A. Tonello, M. Fabert, B. Kibler, T. Sylvestre, A. Barthélémy, V. Couderc, S. Wabnitz, and G. Millot, "Far-detuned cascaded intermodal four-wave mixing in a multimode fiber," Opt. Lett. 42, 1293–1296 (2017).
[10] A. Bendahmane, K. Krupa, A. Tonello, D. Modotto, T. Sylvestre, V. Couderc, S. Wabnitz, and G. Millot, "Seeded intermodal four-wave mixing in a highly multimode fiber," J. Opt. Soc. Am. B 35, 295– 301 (2018).
[11] S. Longhi, "Modulation instability and space-time dynamics in nonlinear parabolic-index optical fibers," Opt. Lett. 28, 2363–2365 (2003).
[12] L. G. Wright, S. Wabnitz, D. N. Christodoulides, and F. W. Wise, "Ultrabroadband dispersive radiation by spatiotemporal oscillation of multimode waves," Phys. Rev. Lett. 115, 223902 (2015).
[13] D. Ceoldo, K. Krupa, A. Tonello, V. Couderc, D. Modotto U. Minoni, G. Millot, and S. Wabnitz, "Second harmonic generation in multimode graded-index fibers: spatial beam cleaning and multiple harmonic sideband generation," Opt. Lett. 42, 971-974 (2017)
[14] Eftekhar, M. A.; Sanjabi-Eznaveh, Z.; Antonio-Lopez, J. E.; Wise, F. W.; Christodoulides, D. N.; Amezcua-Correa, R., "Instant and efficient second-harmonic generation and downconversion in unprepared graded-index multimode fibers," Opt. Lett. 42 (17) 3178-3481 (2017).
[15] K. Krupa, C. Louot, V. Couderc, M. Fabert, R. Guenard, B. M. Shalaby, A. Tonello, D. Pagnoux, P. Leproux, A. Bendahmane, R. Dupiol, G. Millot, and S. Wabnitz, "Spatiotemporal characterization of supercontinuum extending from the visible to the mid-infrared in multimode graded-index optical fiber," Opt. Lett. 41, 5785-5788 (2016).
[16] G. Lopez-Galmiche, Z. S. Eznaveh, M. A. Eftekhar, J. A. Lopez, L. G. Wright, F. Wise, D. Christodoulides, and R. A. Correa, "Visible supercontinuumgeneration in a graded index multimode fiber pumped at 1064 nm," Opt. Lett. 41, 2553–2556 (2016).
[17] R. Guenard, K. Krupa, R. Dupiol, M. Fabert, A. Bendahmane, V. Kermene, A. Desfarges-Berthelemot, J. L. Auguste, A. Tonello, A. Barthélémy, G. Millot, S. Wabnitz, and V. Couderc, "Nonlinear beam self-cleaning in a coupled cavity composite laser based on multimode fiber," Opt. Express 25, 22219–22227 (2017).
[18] L. G. Wright, D. N. Christodoulides, and F. W. Wise, "Spatiotemporal mode-locking in multimode fiber lasers," Science 358, 94–97 (2017).
[19] O. Tzang, A. M. Caravaca-Aguirre, K. Wagner, and R. Piestun, "Adaptive wavefront shaping for controlling nonlinear multimode interactions in optical fibres," Nat. Photonics 12, 368–374 (2018).
[20] E. Deliancourt, M. Fabert, A. Tonello, K. Krupa, A. Desfarges-Berthelemot, V. Kermene, G. Millot, A. Barthélémy, S. Wabnitz, and V. Couderc, "Kerr beam self-cleaning on the LP11 mode in graded-index multimode fiber," OSA Continuum 2(4), 1089-1096 (2019).
[21] E. Deliancourt, M. Fabert, A. Tonello, K. Krupa, A. Desfarges-Berthelemot, V. Kermene, G. Millot , A. Barthélémy, S. Wabnitz,[3] and V. Couderc," Wavefront shaping for optimized many-mode Kerr beam self-cleaning in graded-index multimode fiber," to appear in Opt. Express (2019)
[22] 14. R. Guenard, K. Krupa, R. Dupiol, M. Fabert, A. Bendahmane, V. Kermene, A. Desfarges-Berthelemot, J. L. Auguste, A. Tonello, A. Barthélémy, G. Millot, S. Wabnitz, and V. Couderc, "Kerr self-cleaning of pulsed beam in an ytterbium doped multimode fiber," Opt. Express 25, 4783–4792 (2017).
[23] K. Krupa, V. Couderc, A. Tonello, D. Modotto, A. Barthélémy, G. Millot, S. Wabnitz, "Refractive index profile tailoring of multimode optical fibers for the spatial and spectral shaping of parametric sidebands", J. Opt. Soc. Am. B , vol. 36, 1117-1126 (2019)
[24] Ballato, John; Snitzer, Elias, "Fabrication of fibers with high rare-earth concentrations for Faraday isolator applications", Applied Optics 34(30) 6848-6854 (1995).
[25] Jean-Louis Auguste, Georges Humbert, Stéphanie Leparmentier, Maryna Kudinova, Pierre-Olivier Martin, Gaëlle Delaizir, Kay Schuster and Doris Litzkendorf, "Modified Powder-in-Tube Technique Based on the Consolidation Processing of Powder Materials for Fabricating Specialty Optical Fibers," Materials, 7, 6045-6063 (2014).
[26] Clare, A.G.; Parker, J.M. "The effect of refractive index modifiers on the thermal expansion coefficient of fluoride glasses," Phys. Chem. Glasses, 30, 205–210 (1989).
[27] Doris Litzkendorf, Stephan Grimm, Kay Schuster, Jens Kobelke, Anka Schwuchow, Anne Ludwig, Johannes Kirchhof, Martin Leich, Sylvia Jetschke, and Jan Dellith, Jean-Louis Auguste and Georges Humbert, "Study of Lanthanum Aluminum Silicate Glasses for Passive and Active Optical Fibers," International Journal of Applied Glass Science, 3, [4], 321–331 (2012).
[28] Lemaire, P.J., "Reliability of optical fibers exposed to hydrogen: Prediction of long-term loss increases. Opt. Eng. 1991, 30, 780–790.
[29] Stephanie Leparmentier, "Développement et caractérisation de fibres optiques multimatériaux verre/silice ou verre/air/silice réalisées par un

procédé basé sur l'utilisation de poudre de verres", Thèse n° 98-2010 University of Limoges.

[30] K. Krupa, A. Tonello, V. Couderc, A. Barthelemy, G. Millot, D. Modotto, and S. Wabnitz, "Spatiotemporal light beam compression from nonlinear mode coupling", Physical Review A, vol. 97, 043836 (2018).